\documentclass[aps,prb,twocolumn,floatfix]{revtex4}
\usepackage{bm}
\usepackage{epsf}
\usepackage{amssymb}
\usepackage{amsmath}
\usepackage{graphicx}
\usepackage{rotating}
\usepackage{epsfig}
\usepackage{psfrag}
\usepackage{amsmath}
\usepackage{hyperref}
\usepackage{subfigure}
\usepackage{braket}
\usepackage{tikz}
\usepackage{stmaryrd}
\usepackage{wasysym}
\usepackage{xcolor}

\begin{document}
\title{Moir\'e-mediated phases in synthetic Kondo superlattices}

\author{Jyotirish Das, Onur Erten}
\affiliation{Department of Physics, Arizona State University, Tempe, AZ 85287, USA }

\begin{abstract}
Motivated by the recent experiments on van der Waals heterostructures involving metallic and Mott insulating layers, we construct a moir\'e extension of the Kondo-Heisenberg model and study its phase diagram via Abrikosov fermion mean field theory in one and two dimensions. Our analysis indicates that the stacking dependent Kondo interaction can lead to a variety of new phases. In particular, we show that magnetic order and heavy quasiparticles can nucleate at different regions in the moir\'e unit cell leading to a macroscopic real space phase separation. We observe that these factors can lead to a metal to Kondo insulator percolation transition in two dimensions. Our results highlight the rich physics that can arise in moir\'e superlattices that are composed of inherently strongly correlated layers.
\end{abstract}
\maketitle

\section{Introduction}
The Kondo model describes the interaction among localized magnetic moments and weakly-interacting conduction electrons. It is commonly utilized to analyze a diverse set of many-body phenomena such as screening of magnetic moments in magnetic alloys and quantum dots, heavy fermion formation and unconventional superconductivity.\cite{10.1143/PTP.32.37, Coleman_2015,RevModPhys.56.755,Wirth2016}. In general, the Kondo model is applied to f-electron intermetallics where the local moments and the conduction electrons reside at the same lanthanide/actinide sites. Therefore, mainly an onsite (local) Kondo interaction is considered to describe these systems.

Recently, an alternative direction to form Kondo lattices has been proposed in van der Waals (vdW) heterostructures. These systems separate the two essential degrees of freedom, local moments and conduction electrons, to two different layers. One layer is a Mott insulator with local moments and the other layer is a simple metal or a semimetal such as graphene. The virtual hopping processes between the layers can create an effective Kondo interaction. However, in general these heterobilayers have a lattice mismatch which can give rise to a moir\'e pattern even in the absence of twisting. One such example is the superlattice composed of 2H-TaS$_2$ which is metallic and 1T-TaS$_2$ which is a Mott insulator\cite{https://doi.org/10.1002/smll.202303275}. Since 1T-TaS$_2$ (2H-TaS$_2$) has a lattice constant of 3.36 (3.316)\r{A} \cite{PhysRevB.8.3719,FClerc_2004}, this heterostructure results in moir\'e superlattice with a periodicity of $L \sim$  75 unit cells. Recent scanning tunneling spectroscopy experiments show heavy quasiparticles in 1T/2H TaS$_2$ heterostructures with a coherence temperature of T$_K \sim$ 27 K. Similar behavior demonstrating Kondo screening has also been observed in 1T/1H-TaSe$_2$\cite{Wan_NatComm2023},  1T/1H-TaS$_2$\cite{Vano_Nature2021} and 1T/2H-NbSe$_2$\cite{Liu_SciAdv2021} heterobilayers. Apart from the TaS$_2$ and TaSe$_2$ superlattices, synthetic Kondo lattices may also form by heterostructuring two-dimensional (2D) magnets\cite{Blei_APR2021} with metals or semimetals. For instance, a Kondo model has been proposed for $\alpha$-RuCl$_3$/graphene bilayers\cite{Jin_PRB2021}. Recent studies suggest that the interlayer interaction significantly alters the electronic and magnetic properties of both $\alpha$-RuCl$_3$ and graphene layers\cite{zhou2019, biswas2019, leeb_PRL2021}. 

Moir\'e superlattices of vdW materials have been proven to be tunable quantum platforms for realization of emergent phases\cite{He_ACSNano2021}. While moir\'e engineering of electronic phases in semiconductors and graphene has been extensively studied, the investigation into moir\'e superlattices composed of Mott insulators and magnets is at its early stages. Presently, only a limited number of theoretical investigations have been undertaken\cite{Hejazi2020, Akram_PRB2021, Hejazi_PRB2021, Akram_NanoLett2021, Nica_npjQM2023, Akram_NanoLett2024} and several of these predicted phases have been demonstrated experimentally\cite{Xu_NatNano2022, Song_Science2021, Xie_NatPhys2023}. Regarding moir\'e Kondo lattices, a recent study shows that the stacking dependent Kondo interaction can give rise to complex multi-domain magnetic textures\cite{Keskiner_NanoLett2024}.

\begin{figure}[t!]
    \includegraphics[width=1\linewidth]{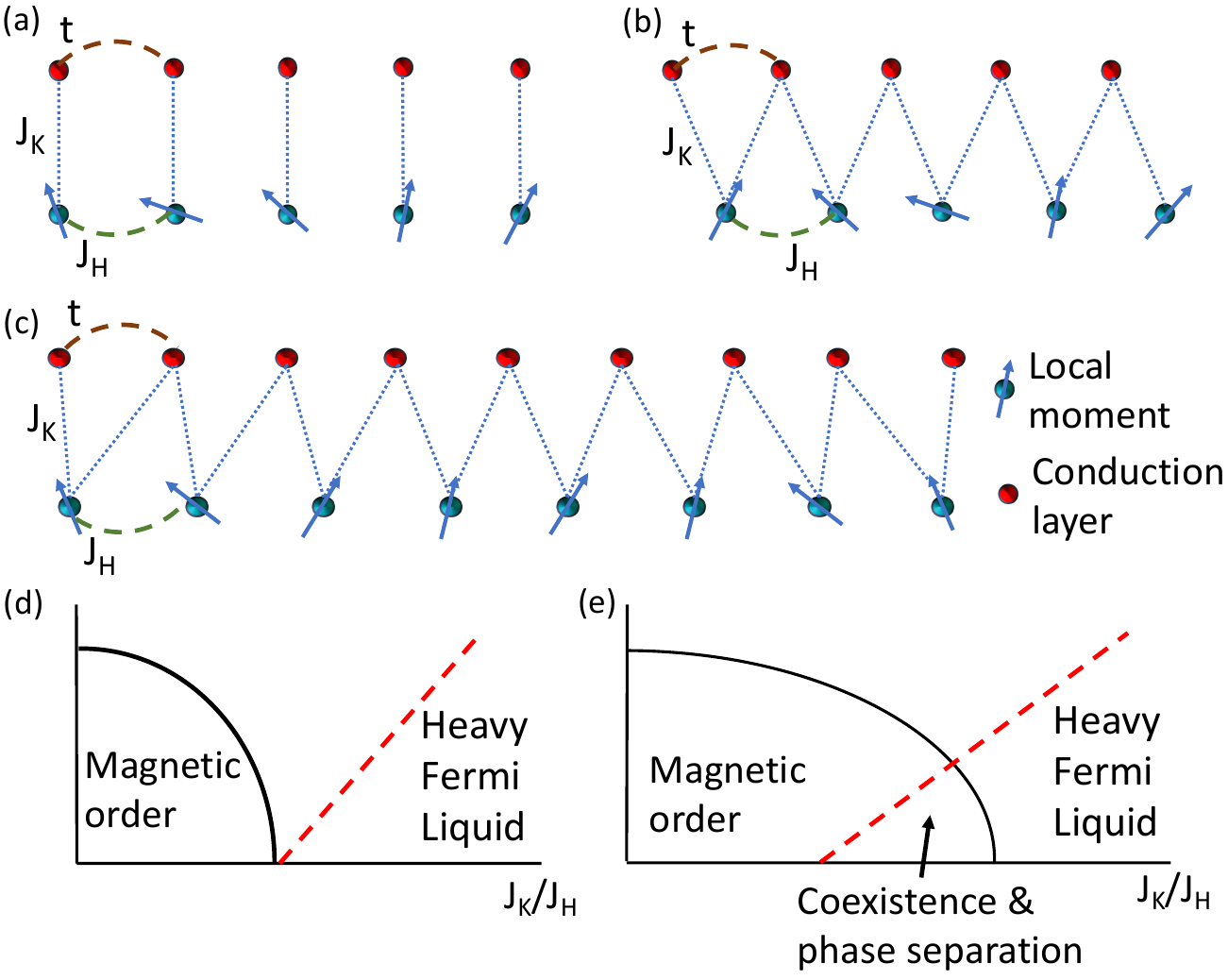}
    \caption{1D schematic of our model: (a) local and (b) staggered stacking pattern in the absence of a lattice mismatch. (c) 1D moir\'e superlattice formed by lattice mismatch between the layers. (d) Schematics of (d) Doniach phase diagram of Kondo lattice model and (e) its moir\'e counterpart. The moir\'e Kondo lattice exhibits a coexistence region where magnetic order and heavy Fermi liquid phases reside at different regions within moir\'e unit cell.}
    \label{Fig1}
\end{figure}

These recent experimental developments motivate us to propose a moir\'e extension to the Kondo-Heisenberg model where the Kondo coupling depends on the distance between the conduction electrons and the local moments (see Fig.~\ref{Fig1}(c)). This extended Kondo interaction is crucial since there are only vanishing number of sites within a moir\'e unit cell that are exactly aligned on top of each other. We introduce a lattice mismatch between the layers which gives rise to a moir\'e pattern in both 1D and 2D. Our main results are summarized as follows: (i) when the two layers have the same lattice constant (no moir\'e pattern), the phase diagram depends on how the two layers are aligned (stacking pattern). In particular, we show that the staggered stacking pattern (see Fig.~\ref{Fig1}(b)) has a higher tendency to order magnetically compared to local stacking (Fig.~\ref{Fig1}(a)). (ii) Unlike the standard Kondo-Heisenberg model, the ground state of its moir\'e counterpart shows a coexistence of magnetically ordered and heavy fermion phases within the moir\'e unit cell in both 1D and 2D for a wide range of parameters. (iii) The different phases in the moir\'e Kondo-Heisenberg model cannot be explained simply by the spatial variation of the Kondo coupling. Instead the local stacking order determines the phase of certain regions within the moir\'e unit cell. (iv) 2D phase diagrams show resemblance to 1D where the heavy Fermions and magnetic order occupy different regions of the moir\'e unit cell. Our analysis indicates that a percolation transition from a metal to a Kondo insulator can take place as a function of model parameters.

Rest of the article is organized as follows. In Section II, we introduce the moir\'e extension of the Kondo-Heisenberg model and discuss the Abrikosov fermion mean theory. In Section III, we first discuss the phase diagram for different stacking orders in the absence of a moir\'e pattern in 1D. Next, we present the phase diagram of moir\'e superlattices and argue that its properties can be understood as a superposition of different stacking patterns. We investigate the effects of varying moir\'e periodicity, and chemical potential. We present the phase diagram in 2D and conclude with a summary and outlook.

\section{Model and methods}
We consider the moir\'e extension of the Kondo-Heisenberg model that describes the interaction between the local moments and conduction electrons at different layers (see Fig.~\ref{Fig1}(c)).
\begin{eqnarray}
    H&=& -t\sum_{\langle ij \rangle;\sigma} c_{i\sigma}^\dagger c_{j\sigma}-\mu \sum_{i \sigma} c_{i\sigma}^\dagger c_{i\sigma}+ \sum_{ij} J_K(r_{ij}) \; \mathbf{S_i} \cdot \boldsymbol{\sigma_j} \nonumber \\ &-& \sum_{\langle ij \rangle} J_H \; \mathbf{S_i \cdot S_j}  \, .
    \label{kh}
\end{eqnarray}
where $c^\dagger_{\sigma}$ denotes the conduction electron creation operator and $\mathbf{S}$ and $\boldsymbol{\sigma}$ denote the spin operators of local moments and conduction electrons respectively. $r_{ij}$ is the separation between the sites $i$ and $j$, $r_{ij}=|{\bf r}_i-{\bf r}_j|$. We consider a stacking dependent extended Kondo interaction given by
\begin{eqnarray}
J_K(r_{ij}) &=& J_K^0 e^{- \alpha r_{ij}/a} ~ {\rm for}~r_{ij}<a \nonumber \\
&=&0~{\rm otherwise}
\label{jk}   
\end{eqnarray}
where $a$ is the lattice constant of conduction electron layer and $\alpha$ is the decay constant which we consider to be between $\alpha\sim 1.1-7$.\cite{Keskiner_NanoLett2024} Note that even though a local Kondo interaction is commonly used for f-electron systems, extended Kondo interaction is essential for topological and various broken symmetry phases in heavy fermions\cite{Alexandrov_PRL2015, Ahamed_PRB2018, Ghazaryan_NJP2021, Vijayvargia_PRB2024}. For the remainder of our article, we consider ferromagnetic Heisenberg interaction, $J_H>0$ and $t=1$. 

In one dimension, a lattice mismatch between the layers can be defined as $\delta = 1 - N_f/N_c$ where $N_f$ ($N_c$) are the number of local moment (conduction electron) sites within a single moir\'e unit cell. The lattice mismatch gives rise to a moir\'e periodicity of $L = 1/\delta$. Note that for such a moir\'e superlattice, $r_{ij}=0$ at the ends of the unit cell (local stacking) whereas $r_{ij}=a/2$ at the center (staggered stacking) as shown in Fig.~\ref{Fig1}. The periodicity in $r_{ij}$ gives rise to a periodicity in the Kondo coupling strength (eq.~\ref{jk}) which in turn leads to a periodicity in observable quantities such as magnetization. In 2D, we consider square lattice for both layers and a lattice mismatch between the layers leads to a moir\'e pattern, similar to 1D. 

\begin{figure}[t!]
\includegraphics[width=1\linewidth]{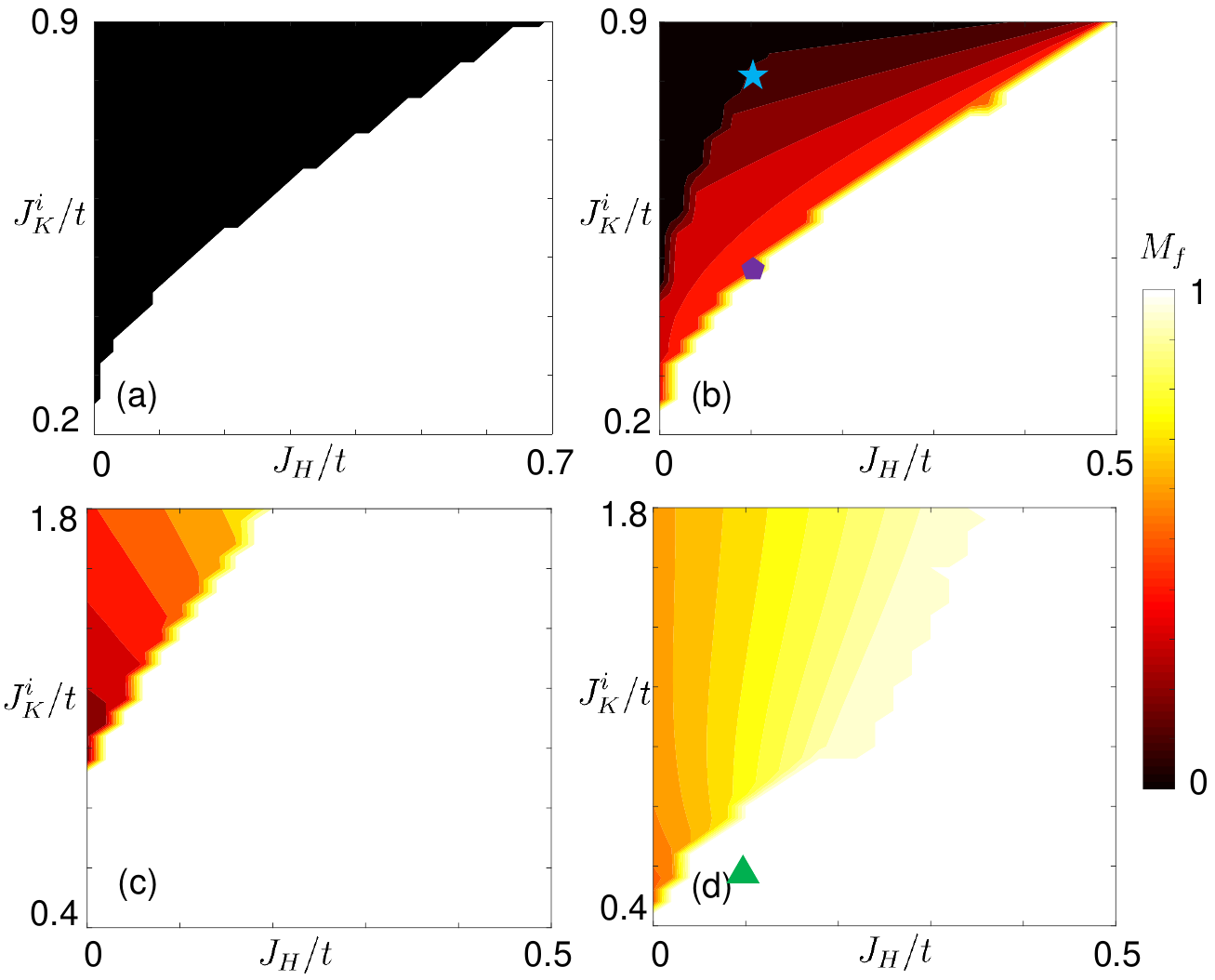}
\caption{Ground state phase diagram in 1D for different stacking orders, without a lattice mismatch (no moir\'e pattern). The color plot denotes the local moment magnetization ($M_f$) as a function of $J_H$ and $J_K$ for (a) local stacking, $\mu=0$; (b) local stacking, $\mu =-1$; (c) staggered stacking, $\mu =0$ and (d) staggered stacking, $\mu =-1$. The star, triangle and pentagon markers indicate certain parameters in Fig.~\ref{Fig3}.}
    \label{Fig2}
\end{figure}

Next, we represent the local moments via Abrikosov fermions, $\mathbf{S}_i= f_{i \alpha}^\dagger \mathbf{\boldsymbol{\sigma}_{\alpha \beta}}f_{i \beta}$ and decouple the Heisenberg exchange within mean field theory as follows, $\sum_{\langle ij \rangle} J_H \; \mathbf{S_i \cdot S_j} \simeq J_H \sum_{i,\nu} S_i^z \langle S_{i+\nu}^z \rangle + {\rm constant \; terms} $ where we picked $z$ as the axis of quantization for the magnetic order and $\nu$ denotes the nearest neighbors. We also decouple the Kondo interaction in magnetic and hybridization channels\cite{}, $\sum_{\langle i j \rangle} J_K(r_{ij}) \mathbf{S_i} \cdot \boldsymbol{\sigma}_j \simeq \sum_{i j} - 2 J_K(r_{ij}) \{ \sum_{\sigma} (V_{ij} c_{j \sigma}^{\dagger} f_{i \sigma} + V_{ij}^{*} f_{i \sigma}^{\dagger} c_{j \sigma})-V_{ij} V_{ij}^* \}+ \sum_{ij} J_K(r_{ij}) (S_{i}^z \langle \sigma_{j}^z \rangle + \sigma_{j}^z \langle S_{i}^z \rangle -\langle S_{i}^z \rangle \langle \sigma_{j}^z \rangle)\, $. 

Here $V_{ij} = \sum_{\sigma} \langle f_{i \sigma}^\dagger c_{j \sigma} \rangle$ and $V_{ij}^*$ denotes its complex conjugate. The hybridization channel has been simplified by using the identity $(f_{i \alpha}^\dagger \mathbf{\boldsymbol{\sigma}_{\alpha \beta}}f_{i \beta})(c_{i \alpha'}^\dagger \mathbf{\boldsymbol{\sigma}_{\alpha' \beta'}}f_{i \beta'}) = - 2 (f_{i \alpha}^\dagger c_{i \alpha})(c_{i \beta}^\dagger f_{i \beta})$. A constraint equation is added to the mean field Hamiltonian which restricts the total f occupancy $n_f=1$ on average. We define the effective Kondo interaction for a local moment at site $i$ as $J_K^i = \sum_{\nu}J_K(r_{i\nu})$ and the effective hybridization as $V_i = \sum_{\nu} V_{i\nu}$.

\section{Results and Discussion}
\subsection{Phase diagram for local and staggered stacking patterns}

Before discussing the moir\'e superlattices, we first present the ground state phase diagram in 1D in the absence of a lattice mismatch (no moir\'e pattern) for local and staggered stacking patterns in Fig.~\ref{Fig2}. For local stacking and $\mu=0$ (Fig.~\ref{Fig2}(a)), the ground state is either a Kondo insulator with no magnetization ($M_f=M_c=0$) for large $J_K$ or a fully-polarized magnetic order ($M_f=1$) with no hybridization ($V_{i}=0$) for large $J_H$. Our analysis indicates that there is no coexistence of magnetic order and heavy fermions. On the contrary, Fig.~\ref{Fig2}(b) shows that a uniform coexistence with finite $M$ and $V_i$ can be stabilized for local stacking for $\mu=-1$, particularly when Kondo and Heisenberg coupling constants are of similar strength and large. The key difference between these two cases, $\mu=0$ and $\mu=-1$ is the absence or the presence a Fermi surface which suppresses or enhances the magnetic susceptibility. The local stacking pattern is simply a standard Kondo model with local interactions and our results are in agreement with previous Abrikosov fermion mean field theory calculations\cite{Senthil_PRB2004, Guerci_SciAdv2023, Vijayvargia_PRB2024}.

Compared to the local stacking, magnetic order occupies a much larger portion of the phase diagram for the staggered stacking pattern as shown in Fig.~\ref{Fig2}(c) and (d). This arises due to an additional $\cos(ka/2)$ factor in the mean field equations for the hybridization channel due to the staggered stacking (details of the mean field theory are given in Appendix A). Averaged over the Fermi surface, the additional phase factor further suppresses the heavy fermion formation and therefore promotes the magnetic order. This effect plays a key role in the analysis of the phase diagram of moir\'e superlattices in the next subsection.

\begin{figure}[t!]
    \includegraphics[width=1\linewidth]{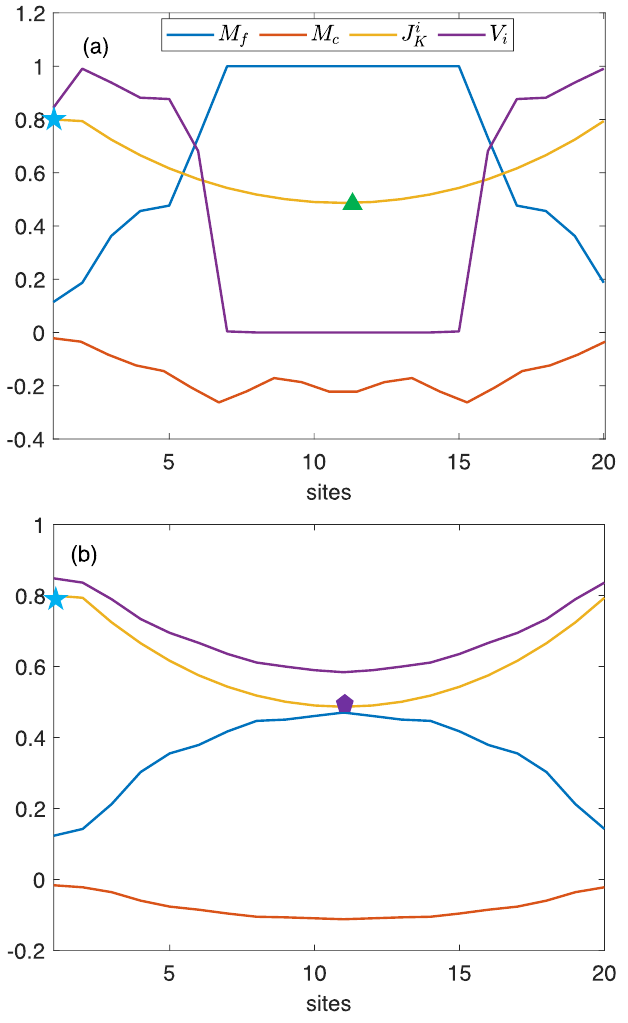}
    \caption{1D self-consistent mean-field solutions (a) a moir\'e superlattice with $N_c = 21$ and $N_f =20$ ($\delta=1/21$) and (b) a local stacking pattern ($\delta = 0$) with an effective Kondo interaction ($J_K^i$) that is periodic and identical to (a). $J_K^0=0.8$ , $\alpha=3$, $\mu=-1$ and $J_H=0.1$ for both panels. Even though the Kondo interaction varies in space the same way in both cases, the magnetic order is much more pronounced in moir\'e superlattices, compared to local stacking. The triangle, star and pentagon markers represent the corresponding $J_K^i$ and $J_H$ values for local and staggered stacking in Fig.~\ref{Fig2}.}
    \label{Fig3}
\end{figure}
\subsection{Phase diagram for 1D moir\'e superlattices}
We present the self-consistent mean-field solutions for 1D moir\'e superlattices in Fig.~\ref{Fig3}(a). We consider a moir\'e unit cell that has $N_c = 21$ and $N_f=20$ sites $(\delta = 1/21)$. We solve this system self-consistently in real space for 5 unit cells with periodic boundary conditions (in total 105 conduction electron sites and 100 local moment sites). The effective Kondo coupling $J_K^i$ is maximum at the ends of the moir\'e unit cell when $r_{ij}=0$ and minimum at the center for $r_{ij}=a/2$. However, the Heisenberg interaction is constant throughout. As a result, the hybridization, $V_i$ is large for the local stacking region whereas magnetic order dominates the center region ($M_f = 1$, $V_i=0$). Due to antiferromagnetic Kondo interaction, magnetic order of the local moment induces a finite and opposite magnetization for conduction electrons ($M_c$). Our results illustrate that this type of phase separation is quite common for a wide range of parameters which we discuss in detail below. 

In order to determine whether the phase diagram of the moir\'e superlattices is completely determined by the spatially varying Kondo interactions, we constructed a model with local stacking ($\delta=0$) for which the Kondo interaction varies within the unit cell, the same way as in the moir\'e superlattice. Our self-consistent mean-field analysis in Fig.~\ref{Fig3}(b) indicates that even though the overall profile of the order parameters is similar, the suppression of the hybridization at the center region of the moir\'e unit cell is highly underestimated in this case. This implies that the ground state of the moir\'e unit cell is not simply controlled by a spatially varying Kondo interaction but the local stacking is crucial as well. For instance, considering the two major stacking patterns within the moir\'e unit cell, local ($r_{ij}=0$) and staggered ($r_{ij}=a/2$) stackings, the order parameters in these regions such as $V_i$ and $M_f$ closely follow non-moir\'e solutions for the corresponding $J_K^i$ as marked by `star' and `triangle' markers in Fig.~\ref{Fig2} and Fig.~\ref{Fig3}. In particular, regarding the center region of the moir\'e unit cell, it is important to take into account the effects of staggered stacking as discussed in the previous subsection. 

\begin{figure}[t]
\includegraphics[width=1\linewidth]{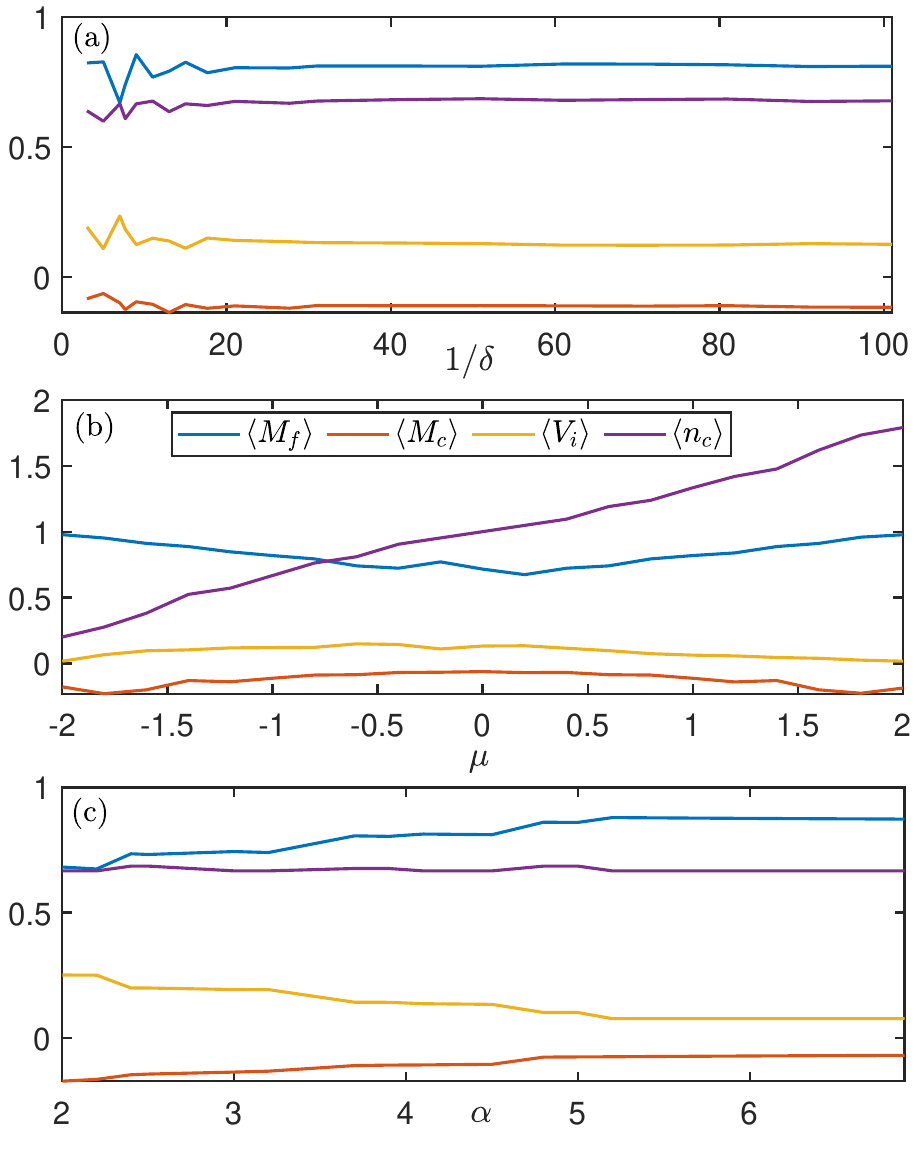}
    \caption{Self-consistent solutions for 1D moir\'e superlattices: the average mean field parameters $\langle M_f \rangle$, $\langle M_c \rangle$, $\langle V_{i} \rangle$ and conduction electron density, $n_c$, as a function of (a) the reciprocal of lattice mismatch ($1/\delta$) for  $\mu=-1$ and $\alpha = 3.7$ (b) the chemical potential ($\mu$) for $1/\delta = 21$ and $\alpha = 4$ and (c) decay constant ($\alpha$) for the extended Kondo interaction for $\mu=-1$ and $1/\delta = 21$. $J_H = 0.1$, $J^0_K = 0.8$ for all panels. }
    \label{Fig4}
\end{figure}

To investigate how these results depend on Hamiltonian parameters, next we discuss the order parameters averaged over the moir\'e unit cell in Fig.~\ref{Fig4}. Fig.~\ref{Fig4}(a) shows that for large moir\'e periodicity ($1/\delta > 20$), the self-consistent solutions does not significantly depend on $L$. However, for small moir\'e periodicity ($1/\delta < 20$), we observe significant fluctuations for the averages of the order parameters as a function of $L$. In this regime, the order parameters change rapidly and therefore the domain wall energies become more important, leading to these fluctuations. Fig.~\ref{Fig4}(b) shows the dependence on the chemical potential for $\delta=1/21$. We find that at half-filling ($\mu=0$), the average hybridization $\langle V_i \rangle$ is strongest since a Kondo insulator can be stabilized. However, the average magnetization is still finite, indicating that the staggered stacking region is still polarized. Deviating from half-filling enlarges the magnetically ordered region. In Appendix B, we present the real space profiles of the order parameters for $\mu=0$ and $\mu=-2$ that shows this effect.

Fig.\ref{Fig4}(c) shows the variation of average order parameters with respect to the decay constant $\alpha$. For large $\alpha$, the Kondo interaction decays faster and therefore the effective Kondo interaction $J_K^i$ becomes smaller within the moir\'e unit cell, particularly away from the local stacking region. As a result, we observe that the average magnetization increases with $\alpha$ and the hybridization decreases (see Appendix B for real space profiles).

\subsection{Phase diagram for 2D moir\'e superlattices}
\begin{figure}[t]
    \includegraphics[width=1\linewidth]{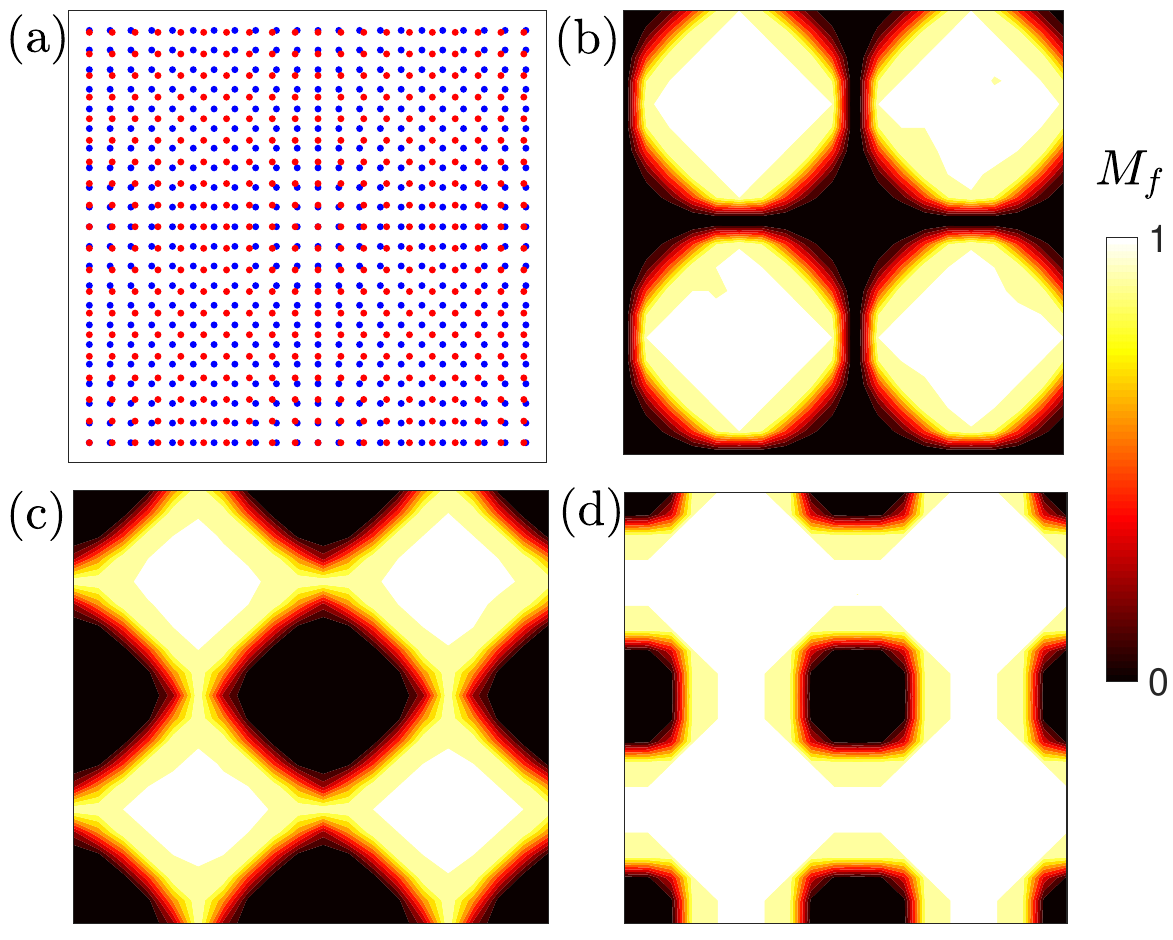}
    \caption{(a) Schematic of a 2D moir\'e unit cell composed of $22 \times 22$ $N_c$ sites (red) and $20 \times 20$ $N_f$ sites (blue). The spatial variation of $M_f$ for $\alpha$ equals to (b) 1.1 (c) 2 and (d) 4. $J_H = 0.02$ and $J_K^0=1.3$ for all panels.}
    \label{Fig5}
\end{figure}
In this subsection, we discuss the mean field phase diagram of the moir\'e extension of the Kondo-Heisenberg model in 2D. We consider 2D moir\'e superlattices that form by two layers of square lattice with a lattice mismatch shown in Fig.~\ref{Fig5}(a). The extended Kondo interaction is still given by eq.~\ref{jk}. Unlike 1D, the local moments in general have 4 conduction electron sites that they interact with via the Kondo coupling. Besides these differences, the general trend in 2D remains quite similar to our analysis in 1D. The heavy fermions nucleate at the local stacking regions whereas the magnetic order occupies the remainder of the moir\'e unit cell. In Fig.~\ref{Fig5}(b)-(d), we show that the ratios of these regions can be controlled by $\alpha$. We consider $\mu=0$ for which the hybridized phase is a Kondo insulator and the polarized phase is a metal. Similar to, 1D magnetically ordered phase grows with $\alpha$. For small $\alpha$, the magnetically ordered metallic regions are separated from each other by a Kondo insulator as shown in Fig.~\ref{Fig5}(b). Since the metallic regions are not connected, the system is an insulator. By increasing $\alpha$ the metallic regions grow and connect with each other which leads to a metal-insulator transition. We thus see an example of a percolation transition which is driven by the moir\'e potential instead of disorder.

We note that our model is oversimplified to realistically model the details of the recent experiments on transition metal dichalcogenide heterobilayers. First of all, the 1T Mott insulating phases in these materials arise due to a ($\sqrt{13} \times \sqrt{13}$) reconstruction of the lattice where the Mott insulator carries effectively one local moment per 13 unit cells of the undistorted lattice. Additionally, a realistic modeling would require an {\it ab initio} band structure of the metallic layer and a symmetry-based matrix element estimation between the layers. However, we believe that the main conclusions we draw from our analysis including the prediction of a macroscopic phase separation may be important for the analysis of the experimental results. We also note that our model is entirely different than the Kondo model realized in MoTe$_2$/WSe$_2$ moir\'e superlattices\cite{Zhao_Nature2023}. MoTe$_2$ and WSe$_2$ are weakly-interacting semiconductors by themselves and the Kondo model emerges only within the moir\'e unit cell. However, our starting point involves Mott insulators which are strongly interacting even as monolayers.

\section{Conclusions}
We introduced a moir\'e extension of the Kondo-Heisenberg model and studied its diagram via self-consistent mean field theory using Abrikosov fermions. Unlike the standard Doniach phase diagram, we found that moir\'e superlattices in general exhibit a real space phase separation where different regions of the moir\'e unit cell order magnetically or form heavy fermions. The properties of these phases are not only determined by the effective Kondo coupling but the local stacking arrangement also plays a key role. A staggered stacking leads to a greater overall magnetization compared to local stacking for the same effective value of the Kondo coupling. We showed that this behavior broadly holds true for 2D as well. Lastly, we highlighted a special case in 2D where we observed a metal-insulator percolation transition as a function of decay constant of the Kondo interaction. Interesting future directions include the determining the properties of domain walls between the magnetic and hybridized phases and a realistic modeling of 1T/1H TaS$_2$ and TaSe$_2$ heterostructures, incorporating the stacking dependent Kondo interaction. 

\section{Acknowledgements}
We thank Pouyan Ghaemi and Tamaghna Hazra for fruitful discussions. This work is supported by NSF Award No. DMR 2234352. 
\appendix

\section{Details of the mean-fields calculations for local and staggered stacking patterns}
For local stacking, the Hamiltonian is
\begin{eqnarray}
    H&=& \sum_{k \sigma} (\epsilon_k -\mu_c) \; c_{k \sigma}^\dagger c_{k \sigma} + J_K \sum_{i}  \; \mathbf{S}_i \cdot \boldsymbol{\sigma}_i
    \nonumber \\ &-& \sum_{\langle ij \rangle} J_H \; \mathbf{S}_i \cdot \mathbf{S}_j  \, .
    \label{org1}
\end{eqnarray}
Following the procedure described in Section II, the mean field Hamiltonian can be expressed as
\begin{equation}
    H=
    \left( 
    \begin{array}{c c c c}
     c_{k \uparrow}^{\dagger} & c_{k \downarrow}^{\dagger} & f_{k \uparrow}^{\dagger} &
     f_{k \downarrow}^{\dagger}
    \end{array} \right) Q_{\rm local} 
    \left(
    \begin{array}{c}
    c_{k \uparrow} \\
    c_{k \downarrow} \\
    f_{k \uparrow}\\
    f_{k \downarrow}
    \end{array} \right) 
\end{equation}
where
\begin{widetext}
\[
    Q_{\rm local}=
    \left(
    \begin{array}{c c|c c}
      \epsilon_k - \mu_c + J_K M_f & & -2J_K V & \\
      & \epsilon_k - \mu_c - J_K M_f & &-2J_K V \\
      \hline
      -2J_K V^*& & -2 J_H M_f + J_K M_c + \lambda  & \\
      &-2J_K V^*& & 2 J_H M_f - J_K M_c + \lambda
    \end{array}
    \right) 
  \]
\end{widetext}
The Lagrange multiplier, $\lambda$ is included to enforce the constraint $n_f=1$ on average. For the staggered stacking arrangement, the Kondo interaction is non-local and can be decoupled via the hybridization and magnetic channels: 
\begin{eqnarray}
    J_K \; \sum_{ i j } \mathbf{S_i \cdot \sigma_j} &=& - 2 J_K \sum_{i j}   \sum_{\sigma} (V_{ij} c_{j \sigma}^{\dagger} f_{i \sigma} + V_{ij}^{*} f_{i \sigma}^{\dagger} c_{j \sigma}) \nonumber \\
   &+& J_K \sum_{ij}  (S_{iz} \langle \sigma_{jz} \rangle + \sigma_{jz} \langle S_{iz} \rangle) \nonumber \\
   &+& {\rm constant \; terms}.
\end{eqnarray}

The 1st term in the spin channel can be expanded as 
\begin{align}
    &- 2 J_K \sum_{i j \sigma}  V_{ij} c_{j \sigma}^{\dagger} f_{i \sigma} \\
    &= - 2 J_K \sum_{i \sigma} (V_{i, i-\frac{1}{2}} c_{i-\frac{1}{2}, \sigma}^{\dagger} f_{i \sigma}+V_{i, i+\frac{1}{2}} c_{i+\frac{1}{2}, \sigma}^{\dagger} f_{i \sigma} ) \nonumber \\
    &= -2J_K \sum_{k \sigma} c^{\dagger}_{k \sigma} f_{k \sigma} (V_{i, i-\frac{1}{2}} e^{i ka/2} + V_{i, i+\frac{1}{2}} e^{-i ka/2}) \nonumber
\end{align}

\begin{figure}[t]
\includegraphics[width=1\linewidth]{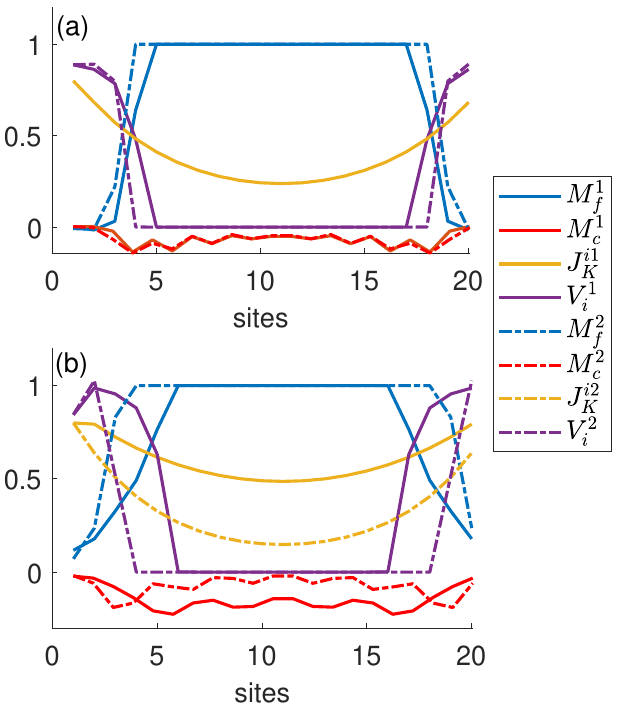}
    \caption{Self consistent solutions for the moir\'e Kondo model in 1D showing the variation with respect to $\mu$ and $\alpha$. (a) $\alpha=4$ and 1(2) corresponds to $\mu=0 (-2)$  and (b) $\mu=-1$ and 1(2) corresponds to $\alpha= 2.5 (5)$. $J_H = 0.1$, $J^0_K = 0.8$ and $1/\delta = 21$ for both the panels.}
    \label{Fig6}
\end{figure}

$i\pm \frac{1}{2}$ simply denote the two nearest c sites for each f site i and $V_{i, i-\frac{1}{2}} = \sum_{\sigma} \langle f^{\dagger}_{i \sigma} c_{i-\frac{1}{2}\sigma} \rangle $.  The mean field Hamiltonian can be written up to a constant term as
\begin{equation}
    H=
    \left( 
    \begin{array}{c c c c}
     c_{k \uparrow}^{\dagger} & c_{k \downarrow}^{\dagger} & f_{k \uparrow}^{\dagger} &
     f_{k \downarrow}^{\dagger}
    \end{array} \right) Q_{\rm staggered} 
    \left(
    \begin{array}{c}
    c_{k \uparrow} \\
    c_{k \downarrow} \\
    f_{k \uparrow}\\
    f_{k \downarrow}
    \end{array} \right) 
\end{equation}
where
\begin{widetext}
\[
    Q_{\rm staggered}=
    \left(
    \begin{array}{c c|c c}
      \epsilon_k - \mu_c + 2 J_K M_f & & -4J_K V {\rm cos}(ka/2) & \\
      & \epsilon_k - \mu_c - 2 J_K M_f & &-4J_K V {\rm cos}(ka/2) \\
      \hline
      -4J_K V^* {\rm cos}(k/2)& & -2 J_H M_f + 2 J_K M_c + \lambda  & \\
      &-4J_K V^* {\rm cos}(k/2)& & 2 J_H M_f - 2 J_K M_c + \lambda
    \end{array}
    \right) 
  \]
\end{widetext}

\section{Real space profiles of the self-consistent solutions for varying $\mu$ and $\alpha$}
In Fig.~\ref{Fig6}(a), we present the self-consistent solutions for 1D moir\'e superlattices with $\mu=0$ (solid lines) and $\mu=-2$ (dashed lines). As discussed in the manuscript, the hybridization channel is strongest at half-filling ($\mu=0$). For $\mu=-2$, the region of the magnetically ordered phase is significantly larger.

In Fig.~\ref{Fig6}(b), we investigate how varying $\alpha$ modifies the solutions. Comparing $\alpha=2.5$ (solid lines) and $\alpha=5$ (dashed lines), it is clear that the effective Kondo coupling, $J_K^i$ is smaller for $\alpha=5$, except the local stacking region. As a result, the magnetic order becomes stronger with increasing $\alpha$.

\bibliographystyle{apsrev4-1}
\bibliography{references.bib}

\begin{thebibliography}{33}%
\makeatletter
\providecommand \@ifxundefined [1]{%
 \@ifx{#1\undefined}
}%
\providecommand \@ifnum [1]{%
 \ifnum #1\expandafter \@firstoftwo
 \else \expandafter \@secondoftwo
 \fi
}%
\providecommand \@ifx [1]{%
 \ifx #1\expandafter \@firstoftwo
 \else \expandafter \@secondoftwo
 \fi
}%
\providecommand \natexlab [1]{#1}%
\providecommand \enquote  [1]{``#1''}%
\providecommand \bibnamefont  [1]{#1}%
\providecommand \bibfnamefont [1]{#1}%
\providecommand \citenamefont [1]{#1}%
\providecommand \href@noop [0]{\@secondoftwo}%
\providecommand \href [0]{\begingroup \@sanitize@url \@href}%
\providecommand \@href[1]{\@@startlink{#1}\@@href}%
\providecommand \@@href[1]{\endgroup#1\@@endlink}%
\providecommand \@sanitize@url [0]{\catcode `\\12\catcode `\$12\catcode `\&12\catcode `\#12\catcode `\^12\catcode `\_12\catcode `\%12\relax}%
\providecommand \@@startlink[1]{}%
\providecommand \@@endlink[0]{}%
\providecommand \url  [0]{\begingroup\@sanitize@url \@url }%
\providecommand \@url [1]{\endgroup\@href {#1}{\urlprefix }}%
\providecommand \urlprefix  [0]{URL }%
\providecommand \Eprint [0]{\href }%
\providecommand \doibase [0]{http://dx.doi.org/}%
\providecommand \selectlanguage [0]{\@gobble}%
\providecommand \bibinfo  [0]{\@secondoftwo}%
\providecommand \bibfield  [0]{\@secondoftwo}%
\providecommand \translation [1]{[#1]}%
\providecommand \BibitemOpen [0]{}%
\providecommand \bibitemStop [0]{}%
\providecommand \bibitemNoStop [0]{.\EOS\space}%
\providecommand \EOS [0]{\spacefactor3000\relax}%
\providecommand \BibitemShut  [1]{\csname bibitem#1\endcsname}%
\let\auto@bib@innerbib\@empty
\bibitem [{\citenamefont {Kondo}(1964)}]{10.1143/PTP.32.37}%
  \BibitemOpen
  \bibfield  {author} {\bibinfo {author} {\bibfnamefont {J.}~\bibnamefont {Kondo}},\ }\href {\doibase 10.1143/PTP.32.37} {\bibfield  {journal} {\bibinfo  {journal} {Progress of Theoretical Physics}\ }\textbf {\bibinfo {volume} {32}},\ \bibinfo {pages} {37} (\bibinfo {year} {1964})},\ \Eprint {http://arxiv.org/abs/https://academic.oup.com/ptp/article-pdf/32/1/37/5193092/32-1-37.pdf} {https://academic.oup.com/ptp/article-pdf/32/1/37/5193092/32-1-37.pdf} \BibitemShut {NoStop}%
\bibitem [{\citenamefont {Coleman}(2015)}]{Coleman_2015}%
  \BibitemOpen
  \bibfield  {author} {\bibinfo {author} {\bibfnamefont {P.}~\bibnamefont {Coleman}},\ }\href@noop {} {\emph {\bibinfo {title} {Introduction to Many-Body Physics}}}\ (\bibinfo  {publisher} {Cambridge University Press},\ \bibinfo {year} {2015})\BibitemShut {NoStop}%
\bibitem [{\citenamefont {Stewart}(1984)}]{RevModPhys.56.755}%
  \BibitemOpen
  \bibfield  {author} {\bibinfo {author} {\bibfnamefont {G.~R.}\ \bibnamefont {Stewart}},\ }\href {\doibase 10.1103/RevModPhys.56.755} {\bibfield  {journal} {\bibinfo  {journal} {Rev. Mod. Phys.}\ }\textbf {\bibinfo {volume} {56}},\ \bibinfo {pages} {755} (\bibinfo {year} {1984})}\BibitemShut {NoStop}%
\bibitem [{\citenamefont {Wirth}\ and\ \citenamefont {Steglich}(2016)}]{Wirth2016}%
  \BibitemOpen
  \bibfield  {author} {\bibinfo {author} {\bibfnamefont {S.}~\bibnamefont {Wirth}}\ and\ \bibinfo {author} {\bibfnamefont {F.}~\bibnamefont {Steglich}},\ }\href {\doibase 10.1038/natrevmats.2016.51} {\bibfield  {journal} {\bibinfo  {journal} {Nature Reviews Materials}\ }\textbf {\bibinfo {volume} {1}},\ \bibinfo {pages} {16051} (\bibinfo {year} {2016})}\BibitemShut {NoStop}%
\bibitem [{\citenamefont {Ayani}\ \emph {et~al.}(2024)\citenamefont {Ayani}, \citenamefont {Pisarra}, \citenamefont {Ibarburu}, \citenamefont {Garnica}, \citenamefont {Miranda}, \citenamefont {Calleja}, \citenamefont {Martín},\ and\ \citenamefont {Vázquez~de Parga}}]{https://doi.org/10.1002/smll.202303275}%
  \BibitemOpen
  \bibfield  {author} {\bibinfo {author} {\bibfnamefont {C.~G.}\ \bibnamefont {Ayani}}, \bibinfo {author} {\bibfnamefont {M.}~\bibnamefont {Pisarra}}, \bibinfo {author} {\bibfnamefont {I.~M.}\ \bibnamefont {Ibarburu}}, \bibinfo {author} {\bibfnamefont {M.}~\bibnamefont {Garnica}}, \bibinfo {author} {\bibfnamefont {R.}~\bibnamefont {Miranda}}, \bibinfo {author} {\bibfnamefont {F.}~\bibnamefont {Calleja}}, \bibinfo {author} {\bibfnamefont {F.}~\bibnamefont {Martín}}, \ and\ \bibinfo {author} {\bibfnamefont {A.~L.}\ \bibnamefont {Vázquez~de Parga}},\ }\href {\doibase https://doi.org/10.1002/smll.202303275} {\bibfield  {journal} {\bibinfo  {journal} {Small}\ }\textbf {\bibinfo {volume} {20}},\ \bibinfo {pages} {2303275} (\bibinfo {year} {2024})}\BibitemShut {NoStop}%
\bibitem [{\citenamefont {Mattheiss}(1973)}]{PhysRevB.8.3719}%
  \BibitemOpen
  \bibfield  {author} {\bibinfo {author} {\bibfnamefont {L.~F.}\ \bibnamefont {Mattheiss}},\ }\href {\doibase 10.1103/PhysRevB.8.3719} {\bibfield  {journal} {\bibinfo  {journal} {Phys. Rev. B}\ }\textbf {\bibinfo {volume} {8}},\ \bibinfo {pages} {3719} (\bibinfo {year} {1973})}\BibitemShut {NoStop}%
\bibitem [{\citenamefont {Clerc}\ \emph {et~al.}(2004)\citenamefont {Clerc}, \citenamefont {Bovet}, \citenamefont {Berger}, \citenamefont {Despont}, \citenamefont {Koitzsch}, \citenamefont {Gallus}, \citenamefont {Patthey}, \citenamefont {Shi}, \citenamefont {Krempasky}, \citenamefont {Garnier},\ and\ \citenamefont {Aebi}}]{FClerc_2004}%
  \BibitemOpen
  \bibfield  {author} {\bibinfo {author} {\bibfnamefont {F.}~\bibnamefont {Clerc}}, \bibinfo {author} {\bibfnamefont {M.}~\bibnamefont {Bovet}}, \bibinfo {author} {\bibfnamefont {H.}~\bibnamefont {Berger}}, \bibinfo {author} {\bibfnamefont {L.}~\bibnamefont {Despont}}, \bibinfo {author} {\bibfnamefont {C.}~\bibnamefont {Koitzsch}}, \bibinfo {author} {\bibfnamefont {O.}~\bibnamefont {Gallus}}, \bibinfo {author} {\bibfnamefont {L.}~\bibnamefont {Patthey}}, \bibinfo {author} {\bibfnamefont {M.}~\bibnamefont {Shi}}, \bibinfo {author} {\bibfnamefont {J.}~\bibnamefont {Krempasky}}, \bibinfo {author} {\bibfnamefont {M.~G.}\ \bibnamefont {Garnier}}, \ and\ \bibinfo {author} {\bibfnamefont {P.}~\bibnamefont {Aebi}},\ }\href {\doibase 10.1088/0953-8984/16/18/026} {\bibfield  {journal} {\bibinfo  {journal} {Journal of Physics: Condensed Matter}\ }\textbf {\bibinfo {volume} {16}},\ \bibinfo {pages} {3271} (\bibinfo {year} {2004})}\BibitemShut {NoStop}%
\bibitem [{\citenamefont {Wan}\ \emph {et~al.}(2023)\citenamefont {Wan}, \citenamefont {Harsh}, \citenamefont {Meninno}, \citenamefont {Dreher}, \citenamefont {Sajan}, \citenamefont {Guo}, \citenamefont {Errea}, \citenamefont {de~Juan},\ and\ \citenamefont {Ugeda}}]{Wan_NatComm2023}%
  \BibitemOpen
  \bibfield  {author} {\bibinfo {author} {\bibfnamefont {W.}~\bibnamefont {Wan}}, \bibinfo {author} {\bibfnamefont {R.}~\bibnamefont {Harsh}}, \bibinfo {author} {\bibfnamefont {A.}~\bibnamefont {Meninno}}, \bibinfo {author} {\bibfnamefont {P.}~\bibnamefont {Dreher}}, \bibinfo {author} {\bibfnamefont {S.}~\bibnamefont {Sajan}}, \bibinfo {author} {\bibfnamefont {H.}~\bibnamefont {Guo}}, \bibinfo {author} {\bibfnamefont {I.}~\bibnamefont {Errea}}, \bibinfo {author} {\bibfnamefont {F.}~\bibnamefont {de~Juan}}, \ and\ \bibinfo {author} {\bibfnamefont {M.~M.}\ \bibnamefont {Ugeda}},\ }\href {\doibase 10.1038/s41467-023-42803-4} {\bibfield  {journal} {\bibinfo  {journal} {Nature Communications}\ }\textbf {\bibinfo {volume} {14}},\ \bibinfo {pages} {7005} (\bibinfo {year} {2023})}\BibitemShut {NoStop}%
\bibitem [{\citenamefont {Va{\v{n}}o}\ \emph {et~al.}(2021)\citenamefont {Va{\v{n}}o}, \citenamefont {Amini}, \citenamefont {Ganguli}, \citenamefont {Chen}, \citenamefont {Lado}, \citenamefont {Kezilebieke},\ and\ \citenamefont {Liljeroth}}]{Vano_Nature2021}%
  \BibitemOpen
  \bibfield  {author} {\bibinfo {author} {\bibfnamefont {V.}~\bibnamefont {Va{\v{n}}o}}, \bibinfo {author} {\bibfnamefont {M.}~\bibnamefont {Amini}}, \bibinfo {author} {\bibfnamefont {S.~C.}\ \bibnamefont {Ganguli}}, \bibinfo {author} {\bibfnamefont {G.}~\bibnamefont {Chen}}, \bibinfo {author} {\bibfnamefont {J.~L.}\ \bibnamefont {Lado}}, \bibinfo {author} {\bibfnamefont {S.}~\bibnamefont {Kezilebieke}}, \ and\ \bibinfo {author} {\bibfnamefont {P.}~\bibnamefont {Liljeroth}},\ }\href {\doibase 10.1038/s41586-021-04021-0} {\bibfield  {journal} {\bibinfo  {journal} {Nature}\ }\textbf {\bibinfo {volume} {599}},\ \bibinfo {pages} {582} (\bibinfo {year} {2021})}\BibitemShut {NoStop}%
\bibitem [{\citenamefont {Liu}\ \emph {et~al.}(2021)\citenamefont {Liu}, \citenamefont {Leveillee}, \citenamefont {Lu}, \citenamefont {Yu}, \citenamefont {Kim}, \citenamefont {Tian}, \citenamefont {Shi}, \citenamefont {Lai}, \citenamefont {Zhang}, \citenamefont {Giustino},\ and\ \citenamefont {Shih}}]{Liu_SciAdv2021}%
  \BibitemOpen
  \bibfield  {author} {\bibinfo {author} {\bibfnamefont {M.}~\bibnamefont {Liu}}, \bibinfo {author} {\bibfnamefont {J.}~\bibnamefont {Leveillee}}, \bibinfo {author} {\bibfnamefont {S.}~\bibnamefont {Lu}}, \bibinfo {author} {\bibfnamefont {J.}~\bibnamefont {Yu}}, \bibinfo {author} {\bibfnamefont {H.}~\bibnamefont {Kim}}, \bibinfo {author} {\bibfnamefont {C.}~\bibnamefont {Tian}}, \bibinfo {author} {\bibfnamefont {Y.}~\bibnamefont {Shi}}, \bibinfo {author} {\bibfnamefont {K.}~\bibnamefont {Lai}}, \bibinfo {author} {\bibfnamefont {C.}~\bibnamefont {Zhang}}, \bibinfo {author} {\bibfnamefont {F.}~\bibnamefont {Giustino}}, \ and\ \bibinfo {author} {\bibfnamefont {C.-K.}\ \bibnamefont {Shih}},\ }\href {\doibase 10.1126/sciadv.abi6339} {\bibfield  {journal} {\bibinfo  {journal} {Science Advances}\ }\textbf {\bibinfo {volume} {7}},\ \bibinfo {pages} {eabi6339} (\bibinfo {year} {2021})}\BibitemShut {NoStop}%
\bibitem [{\citenamefont {Blei}\ \emph {et~al.}(2021)\citenamefont {Blei}, \citenamefont {Lado}, \citenamefont {Song}, \citenamefont {Dey}, \citenamefont {Erten}, \citenamefont {Pardo}, \citenamefont {Comin}, \citenamefont {Tongay},\ and\ \citenamefont {Botana}}]{Blei_APR2021}%
  \BibitemOpen
  \bibfield  {author} {\bibinfo {author} {\bibfnamefont {M.}~\bibnamefont {Blei}}, \bibinfo {author} {\bibfnamefont {J.~L.}\ \bibnamefont {Lado}}, \bibinfo {author} {\bibfnamefont {Q.}~\bibnamefont {Song}}, \bibinfo {author} {\bibfnamefont {D.}~\bibnamefont {Dey}}, \bibinfo {author} {\bibfnamefont {O.}~\bibnamefont {Erten}}, \bibinfo {author} {\bibfnamefont {V.}~\bibnamefont {Pardo}}, \bibinfo {author} {\bibfnamefont {R.}~\bibnamefont {Comin}}, \bibinfo {author} {\bibfnamefont {S.}~\bibnamefont {Tongay}}, \ and\ \bibinfo {author} {\bibfnamefont {A.~S.}\ \bibnamefont {Botana}},\ }\href {\doibase 10.1063/5.0025658} {\bibfield  {journal} {\bibinfo  {journal} {Applied Physics Reviews}\ }\textbf {\bibinfo {volume} {8}},\ \bibinfo {pages} {021301} (\bibinfo {year} {2021})}\BibitemShut {NoStop}%
\bibitem [{\citenamefont {Jin}\ and\ \citenamefont {Knolle}(2021)}]{Jin_PRB2021}%
  \BibitemOpen
  \bibfield  {author} {\bibinfo {author} {\bibfnamefont {H.-K.}\ \bibnamefont {Jin}}\ and\ \bibinfo {author} {\bibfnamefont {J.}~\bibnamefont {Knolle}},\ }\href {\doibase 10.1103/PhysRevB.104.045140} {\bibfield  {journal} {\bibinfo  {journal} {Phys. Rev. B}\ }\textbf {\bibinfo {volume} {104}},\ \bibinfo {pages} {045140} (\bibinfo {year} {2021})}\BibitemShut {NoStop}%
\bibitem [{\citenamefont {Zhou}\ \emph {et~al.}(2019)\citenamefont {Zhou}, \citenamefont {Balgley}, \citenamefont {Lampen-Kelley}, \citenamefont {Yan}, \citenamefont {Mandrus},\ and\ \citenamefont {Henriksen}}]{zhou2019}%
  \BibitemOpen
  \bibfield  {author} {\bibinfo {author} {\bibfnamefont {B.}~\bibnamefont {Zhou}}, \bibinfo {author} {\bibfnamefont {J.}~\bibnamefont {Balgley}}, \bibinfo {author} {\bibfnamefont {P.}~\bibnamefont {Lampen-Kelley}}, \bibinfo {author} {\bibfnamefont {J.-Q.}\ \bibnamefont {Yan}}, \bibinfo {author} {\bibfnamefont {D.~G.}\ \bibnamefont {Mandrus}}, \ and\ \bibinfo {author} {\bibfnamefont {E.~A.}\ \bibnamefont {Henriksen}},\ }\href@noop {} {\bibfield  {journal} {\bibinfo  {journal} {Physical Review B}\ }\textbf {\bibinfo {volume} {100}},\ \bibinfo {pages} {165426} (\bibinfo {year} {2019})}\BibitemShut {NoStop}%
\bibitem [{\citenamefont {Biswas}\ \emph {et~al.}(2019)\citenamefont {Biswas}, \citenamefont {Li}, \citenamefont {Winter}, \citenamefont {Knolle},\ and\ \citenamefont {Valent{\'\i}}}]{biswas2019}%
  \BibitemOpen
  \bibfield  {author} {\bibinfo {author} {\bibfnamefont {S.}~\bibnamefont {Biswas}}, \bibinfo {author} {\bibfnamefont {Y.}~\bibnamefont {Li}}, \bibinfo {author} {\bibfnamefont {S.~M.}\ \bibnamefont {Winter}}, \bibinfo {author} {\bibfnamefont {J.}~\bibnamefont {Knolle}}, \ and\ \bibinfo {author} {\bibfnamefont {R.}~\bibnamefont {Valent{\'\i}}},\ }\href@noop {} {\bibfield  {journal} {\bibinfo  {journal} {Physical Review Letters}\ }\textbf {\bibinfo {volume} {123}},\ \bibinfo {pages} {237201} (\bibinfo {year} {2019})}\BibitemShut {NoStop}%
\bibitem [{\citenamefont {Leeb}\ \emph {et~al.}(2021)\citenamefont {Leeb}, \citenamefont {Polyudov}, \citenamefont {Mashhadi}, \citenamefont {Biswas}, \citenamefont {Valent{\'\i}}, \citenamefont {Burghard},\ and\ \citenamefont {Knolle}}]{leeb_PRL2021}%
  \BibitemOpen
  \bibfield  {author} {\bibinfo {author} {\bibfnamefont {V.}~\bibnamefont {Leeb}}, \bibinfo {author} {\bibfnamefont {K.}~\bibnamefont {Polyudov}}, \bibinfo {author} {\bibfnamefont {S.}~\bibnamefont {Mashhadi}}, \bibinfo {author} {\bibfnamefont {S.}~\bibnamefont {Biswas}}, \bibinfo {author} {\bibfnamefont {R.}~\bibnamefont {Valent{\'\i}}}, \bibinfo {author} {\bibfnamefont {M.}~\bibnamefont {Burghard}}, \ and\ \bibinfo {author} {\bibfnamefont {J.}~\bibnamefont {Knolle}},\ }\href@noop {} {\bibfield  {journal} {\bibinfo  {journal} {Physical Review Letters}\ }\textbf {\bibinfo {volume} {126}},\ \bibinfo {pages} {097201} (\bibinfo {year} {2021})}\BibitemShut {NoStop}%
\bibitem [{\citenamefont {He}\ \emph {et~al.}(2021)\citenamefont {He}, \citenamefont {Zhou}, \citenamefont {Ye}, \citenamefont {Cho}, \citenamefont {Jeong}, \citenamefont {Meng},\ and\ \citenamefont {Wang}}]{He_ACSNano2021}%
  \BibitemOpen
  \bibfield  {author} {\bibinfo {author} {\bibfnamefont {F.}~\bibnamefont {He}}, \bibinfo {author} {\bibfnamefont {Y.}~\bibnamefont {Zhou}}, \bibinfo {author} {\bibfnamefont {Z.}~\bibnamefont {Ye}}, \bibinfo {author} {\bibfnamefont {S.-H.}\ \bibnamefont {Cho}}, \bibinfo {author} {\bibfnamefont {J.}~\bibnamefont {Jeong}}, \bibinfo {author} {\bibfnamefont {X.}~\bibnamefont {Meng}}, \ and\ \bibinfo {author} {\bibfnamefont {Y.}~\bibnamefont {Wang}},\ }\href {\doibase 10.1021/acsnano.0c10435} {\bibfield  {journal} {\bibinfo  {journal} {ACS Nano}\ }\textbf {\bibinfo {volume} {15}},\ \bibinfo {pages} {5944} (\bibinfo {year} {2021})}\BibitemShut {NoStop}%
\bibitem [{\citenamefont {Hejazi}\ \emph {et~al.}(2020)\citenamefont {Hejazi}, \citenamefont {Luo},\ and\ \citenamefont {Balents}}]{Hejazi2020}%
  \BibitemOpen
  \bibfield  {author} {\bibinfo {author} {\bibfnamefont {K.}~\bibnamefont {Hejazi}}, \bibinfo {author} {\bibfnamefont {Z.-X.}\ \bibnamefont {Luo}}, \ and\ \bibinfo {author} {\bibfnamefont {L.}~\bibnamefont {Balents}},\ }\href {\doibase 10.1073/pnas.2000347117} {\bibfield  {journal} {\bibinfo  {journal} {Proceedings of the National Academy of Sciences}\ }\textbf {\bibinfo {volume} {117}},\ \bibinfo {pages} {10721} (\bibinfo {year} {2020})}\BibitemShut {NoStop}%
\bibitem [{\citenamefont {Akram}\ and\ \citenamefont {Erten}(2021)}]{Akram_PRB2021}%
  \BibitemOpen
  \bibfield  {author} {\bibinfo {author} {\bibfnamefont {M.}~\bibnamefont {Akram}}\ and\ \bibinfo {author} {\bibfnamefont {O.}~\bibnamefont {Erten}},\ }\href {\doibase 10.1103/PhysRevB.103.L140406} {\bibfield  {journal} {\bibinfo  {journal} {Phys. Rev. B}\ }\textbf {\bibinfo {volume} {103}},\ \bibinfo {pages} {L140406} (\bibinfo {year} {2021})}\BibitemShut {NoStop}%
\bibitem [{\citenamefont {Hejazi}\ \emph {et~al.}(2021)\citenamefont {Hejazi}, \citenamefont {Luo},\ and\ \citenamefont {Balents}}]{Hejazi_PRB2021}%
  \BibitemOpen
  \bibfield  {author} {\bibinfo {author} {\bibfnamefont {K.}~\bibnamefont {Hejazi}}, \bibinfo {author} {\bibfnamefont {Z.-X.}\ \bibnamefont {Luo}}, \ and\ \bibinfo {author} {\bibfnamefont {L.}~\bibnamefont {Balents}},\ }\href {\doibase 10.1103/PhysRevB.104.L100406} {\bibfield  {journal} {\bibinfo  {journal} {Phys. Rev. B}\ }\textbf {\bibinfo {volume} {104}},\ \bibinfo {pages} {L100406} (\bibinfo {year} {2021})}\BibitemShut {NoStop}%
\bibitem [{\citenamefont {Akram}\ \emph {et~al.}(2021)\citenamefont {Akram}, \citenamefont {LaBollita}, \citenamefont {Dey}, \citenamefont {Kapeghian}, \citenamefont {Erten},\ and\ \citenamefont {Botana}}]{Akram_NanoLett2021}%
  \BibitemOpen
  \bibfield  {author} {\bibinfo {author} {\bibfnamefont {M.}~\bibnamefont {Akram}}, \bibinfo {author} {\bibfnamefont {H.}~\bibnamefont {LaBollita}}, \bibinfo {author} {\bibfnamefont {D.}~\bibnamefont {Dey}}, \bibinfo {author} {\bibfnamefont {J.}~\bibnamefont {Kapeghian}}, \bibinfo {author} {\bibfnamefont {O.}~\bibnamefont {Erten}}, \ and\ \bibinfo {author} {\bibfnamefont {A.~S.}\ \bibnamefont {Botana}},\ }\href {\doibase 10.1021/acs.nanolett.1c02096} {\bibfield  {journal} {\bibinfo  {journal} {Nano Letters}\ }\textbf {\bibinfo {volume} {21}},\ \bibinfo {pages} {6633} (\bibinfo {year} {2021})}\BibitemShut {NoStop}%
\bibitem [{\citenamefont {Nica}\ \emph {et~al.}(2023)\citenamefont {Nica}, \citenamefont {Akram}, \citenamefont {Vijayvargia}, \citenamefont {Moessner},\ and\ \citenamefont {Erten}}]{Nica_npjQM2023}%
  \BibitemOpen
  \bibfield  {author} {\bibinfo {author} {\bibfnamefont {E.}~\bibnamefont {Nica}}, \bibinfo {author} {\bibfnamefont {M.}~\bibnamefont {Akram}}, \bibinfo {author} {\bibfnamefont {A.}~\bibnamefont {Vijayvargia}}, \bibinfo {author} {\bibfnamefont {R.}~\bibnamefont {Moessner}}, \ and\ \bibinfo {author} {\bibfnamefont {O.}~\bibnamefont {Erten}},\ }\href {https://doi.org/10.1038/s41535-023-00541-2} {\bibfield  {journal} {\bibinfo  {journal} {npj Quantum Mater.}\ }\textbf {\bibinfo {volume} {8}},\ \bibinfo {pages} {9} (\bibinfo {year} {2023})}\BibitemShut {NoStop}%
\bibitem [{\citenamefont {Akram}\ \emph {et~al.}(2024)\citenamefont {Akram}, \citenamefont {Kapeghian}, \citenamefont {Das}, \citenamefont {Valent{\'i}}, \citenamefont {Botana},\ and\ \citenamefont {Erten}}]{Akram_NanoLett2024}%
  \BibitemOpen
  \bibfield  {author} {\bibinfo {author} {\bibfnamefont {M.}~\bibnamefont {Akram}}, \bibinfo {author} {\bibfnamefont {J.}~\bibnamefont {Kapeghian}}, \bibinfo {author} {\bibfnamefont {J.}~\bibnamefont {Das}}, \bibinfo {author} {\bibfnamefont {R.}~\bibnamefont {Valent{\'i}}}, \bibinfo {author} {\bibfnamefont {A.~S.}\ \bibnamefont {Botana}}, \ and\ \bibinfo {author} {\bibfnamefont {O.}~\bibnamefont {Erten}},\ }\href {\doibase 10.1021/acs.nanolett.3c04084} {\bibfield  {journal} {\bibinfo  {journal} {Nano Letters}\ }\textbf {\bibinfo {volume} {24}},\ \bibinfo {pages} {890} (\bibinfo {year} {2024})}\BibitemShut {NoStop}%
\bibitem [{\citenamefont {Xu}\ \emph {et~al.}(2022)\citenamefont {Xu}, \citenamefont {Ray}, \citenamefont {Shao}, \citenamefont {Jiang}, \citenamefont {Lee}, \citenamefont {Weber}, \citenamefont {Goldberger}, \citenamefont {Watanabe}, \citenamefont {Taniguchi}, \citenamefont {Muller}, \citenamefont {Mak},\ and\ \citenamefont {Shan}}]{Xu_NatNano2022}%
  \BibitemOpen
  \bibfield  {author} {\bibinfo {author} {\bibfnamefont {Y.}~\bibnamefont {Xu}}, \bibinfo {author} {\bibfnamefont {A.}~\bibnamefont {Ray}}, \bibinfo {author} {\bibfnamefont {Y.-T.}\ \bibnamefont {Shao}}, \bibinfo {author} {\bibfnamefont {S.}~\bibnamefont {Jiang}}, \bibinfo {author} {\bibfnamefont {K.}~\bibnamefont {Lee}}, \bibinfo {author} {\bibfnamefont {D.}~\bibnamefont {Weber}}, \bibinfo {author} {\bibfnamefont {J.~E.}\ \bibnamefont {Goldberger}}, \bibinfo {author} {\bibfnamefont {K.}~\bibnamefont {Watanabe}}, \bibinfo {author} {\bibfnamefont {T.}~\bibnamefont {Taniguchi}}, \bibinfo {author} {\bibfnamefont {D.~A.}\ \bibnamefont {Muller}}, \bibinfo {author} {\bibfnamefont {K.~F.}\ \bibnamefont {Mak}}, \ and\ \bibinfo {author} {\bibfnamefont {J.}~\bibnamefont {Shan}},\ }\href {\doibase 10.1038/s41565-021-01014-y} {\bibfield  {journal} {\bibinfo  {journal} {Nature nanotechnology}\ }\textbf {\bibinfo {volume} {17}},\ \bibinfo {pages} {143—147} (\bibinfo {year} {2022})}\BibitemShut {NoStop}%
\bibitem [{\citenamefont {Song}\ \emph {et~al.}(2021)\citenamefont {Song}, \citenamefont {Sun}, \citenamefont {Anderson}, \citenamefont {Wang}, \citenamefont {Qian}, \citenamefont {Taniguchi}, \citenamefont {Watanabe}, \citenamefont {McGuire}, \citenamefont {St\"ohr}, \citenamefont {Xiao}, \citenamefont {Cao}, \citenamefont {Wrachtrup},\ and\ \citenamefont {Xu}}]{Song_Science2021}%
  \BibitemOpen
  \bibfield  {author} {\bibinfo {author} {\bibfnamefont {T.}~\bibnamefont {Song}}, \bibinfo {author} {\bibfnamefont {Q.-C.}\ \bibnamefont {Sun}}, \bibinfo {author} {\bibfnamefont {E.}~\bibnamefont {Anderson}}, \bibinfo {author} {\bibfnamefont {C.}~\bibnamefont {Wang}}, \bibinfo {author} {\bibfnamefont {J.}~\bibnamefont {Qian}}, \bibinfo {author} {\bibfnamefont {T.}~\bibnamefont {Taniguchi}}, \bibinfo {author} {\bibfnamefont {K.}~\bibnamefont {Watanabe}}, \bibinfo {author} {\bibfnamefont {M.~A.}\ \bibnamefont {McGuire}}, \bibinfo {author} {\bibfnamefont {R.}~\bibnamefont {St\"ohr}}, \bibinfo {author} {\bibfnamefont {D.}~\bibnamefont {Xiao}}, \bibinfo {author} {\bibfnamefont {T.}~\bibnamefont {Cao}}, \bibinfo {author} {\bibfnamefont {J.}~\bibnamefont {Wrachtrup}}, \ and\ \bibinfo {author} {\bibfnamefont {X.}~\bibnamefont {Xu}},\ }\href {\doibase 10.1126/science.abj7478} {\bibfield  {journal} {\bibinfo  {journal} {Science}\ }\textbf {\bibinfo {volume} {374}},\ \bibinfo {pages} {1140} (\bibinfo {year}
  {2021})}\BibitemShut {NoStop}%
\bibitem [{\citenamefont {Xie}\ \emph {et~al.}(2023)\citenamefont {Xie}, \citenamefont {Luo}, \citenamefont {Ye}, \citenamefont {Sun}, \citenamefont {Ye}, \citenamefont {Sung}, \citenamefont {Ge}, \citenamefont {Yan}, \citenamefont {Fu}, \citenamefont {Tian}, \citenamefont {Lei}, \citenamefont {Sun}, \citenamefont {Hovden}, \citenamefont {He},\ and\ \citenamefont {Zhao}}]{Xie_NatPhys2023}%
  \BibitemOpen
  \bibfield  {author} {\bibinfo {author} {\bibfnamefont {H.}~\bibnamefont {Xie}}, \bibinfo {author} {\bibfnamefont {X.}~\bibnamefont {Luo}}, \bibinfo {author} {\bibfnamefont {Z.}~\bibnamefont {Ye}}, \bibinfo {author} {\bibfnamefont {Z.}~\bibnamefont {Sun}}, \bibinfo {author} {\bibfnamefont {G.}~\bibnamefont {Ye}}, \bibinfo {author} {\bibfnamefont {S.}~\bibnamefont {Sung}}, \bibinfo {author} {\bibfnamefont {H.}~\bibnamefont {Ge}}, \bibinfo {author} {\bibfnamefont {S.}~\bibnamefont {Yan}}, \bibinfo {author} {\bibfnamefont {Y.}~\bibnamefont {Fu}}, \bibinfo {author} {\bibfnamefont {S.}~\bibnamefont {Tian}}, \bibinfo {author} {\bibfnamefont {H.}~\bibnamefont {Lei}}, \bibinfo {author} {\bibfnamefont {K.}~\bibnamefont {Sun}}, \bibinfo {author} {\bibfnamefont {R.}~\bibnamefont {Hovden}}, \bibinfo {author} {\bibfnamefont {R.}~\bibnamefont {He}}, \ and\ \bibinfo {author} {\bibfnamefont {L.}~\bibnamefont {Zhao}},\ }\href {\doibase 10.1038/s41567-023-02061-z} {\bibfield  {journal} {\bibinfo  {journal} {Nature Physics}\
  }\textbf {\bibinfo {volume} {19}},\ \bibinfo {pages} {1150} (\bibinfo {year} {2023})}\BibitemShut {NoStop}%
\bibitem [{\citenamefont {A.~Keskiner}\ \emph {et~al.}(2024)\citenamefont {A.~Keskiner}, \citenamefont {Ghaemi}, \citenamefont {Oktel},\ and\ \citenamefont {Erten}}]{Keskiner_NanoLett2024}%
  \BibitemOpen
  \bibfield  {author} {\bibinfo {author} {\bibfnamefont {M.}~\bibnamefont {A.~Keskiner}}, \bibinfo {author} {\bibfnamefont {P.}~\bibnamefont {Ghaemi}}, \bibinfo {author} {\bibfnamefont {M.~{\"O}.}\ \bibnamefont {Oktel}}, \ and\ \bibinfo {author} {\bibfnamefont {O.}~\bibnamefont {Erten}},\ }\href {\doibase 10.1021/acs.nanolett.4c01574} {\bibfield  {journal} {\bibinfo  {journal} {Nano Letters}\ } (\bibinfo {year} {2024}),\ 10.1021/acs.nanolett.4c01574}\BibitemShut {NoStop}%
\bibitem [{\citenamefont {Alexandrov}\ \emph {et~al.}(2015)\citenamefont {Alexandrov}, \citenamefont {Coleman},\ and\ \citenamefont {Erten}}]{Alexandrov_PRL2015}%
  \BibitemOpen
  \bibfield  {author} {\bibinfo {author} {\bibfnamefont {V.}~\bibnamefont {Alexandrov}}, \bibinfo {author} {\bibfnamefont {P.}~\bibnamefont {Coleman}}, \ and\ \bibinfo {author} {\bibfnamefont {O.}~\bibnamefont {Erten}},\ }\href {\doibase 10.1103/PhysRevLett.114.177202} {\bibfield  {journal} {\bibinfo  {journal} {Phys. Rev. Lett.}\ }\textbf {\bibinfo {volume} {114}},\ \bibinfo {pages} {177202} (\bibinfo {year} {2015})}\BibitemShut {NoStop}%
\bibitem [{\citenamefont {Ahamed}\ \emph {et~al.}(2018)\citenamefont {Ahamed}, \citenamefont {Moessner},\ and\ \citenamefont {Erten}}]{Ahamed_PRB2018}%
  \BibitemOpen
  \bibfield  {author} {\bibinfo {author} {\bibfnamefont {S.}~\bibnamefont {Ahamed}}, \bibinfo {author} {\bibfnamefont {R.}~\bibnamefont {Moessner}}, \ and\ \bibinfo {author} {\bibfnamefont {O.}~\bibnamefont {Erten}},\ }\href {\doibase 10.1103/PhysRevB.98.054420} {\bibfield  {journal} {\bibinfo  {journal} {Phys. Rev. B}\ }\textbf {\bibinfo {volume} {98}},\ \bibinfo {pages} {054420} (\bibinfo {year} {2018})}\BibitemShut {NoStop}%
\bibitem [{\citenamefont {Ghazaryan}\ \emph {et~al.}(2021)\citenamefont {Ghazaryan}, \citenamefont {Nica}, \citenamefont {Erten},\ and\ \citenamefont {Ghaemi}}]{Ghazaryan_NJP2021}%
  \BibitemOpen
  \bibfield  {author} {\bibinfo {author} {\bibfnamefont {A.}~\bibnamefont {Ghazaryan}}, \bibinfo {author} {\bibfnamefont {E.~M.}\ \bibnamefont {Nica}}, \bibinfo {author} {\bibfnamefont {O.}~\bibnamefont {Erten}}, \ and\ \bibinfo {author} {\bibfnamefont {P.}~\bibnamefont {Ghaemi}},\ }\href {\doibase 10.1088/1367-2630/ac4124} {\bibfield  {journal} {\bibinfo  {journal} {New Journal of Physics}\ }\textbf {\bibinfo {volume} {23}},\ \bibinfo {pages} {123042} (\bibinfo {year} {2021})}\BibitemShut {NoStop}%
\bibitem [{\citenamefont {Vijayvargia}\ and\ \citenamefont {Erten}(2024)}]{Vijayvargia_PRB2024}%
  \BibitemOpen
  \bibfield  {author} {\bibinfo {author} {\bibfnamefont {A.}~\bibnamefont {Vijayvargia}}\ and\ \bibinfo {author} {\bibfnamefont {O.}~\bibnamefont {Erten}},\ }\href {\doibase 10.1103/PhysRevB.109.L201118} {\bibfield  {journal} {\bibinfo  {journal} {Phys. Rev. B}\ }\textbf {\bibinfo {volume} {109}},\ \bibinfo {pages} {L201118} (\bibinfo {year} {2024})}\BibitemShut {NoStop}%
\bibitem [{\citenamefont {Senthil}\ \emph {et~al.}(2004)\citenamefont {Senthil}, \citenamefont {Vojta},\ and\ \citenamefont {Sachdev}}]{Senthil_PRB2004}%
  \BibitemOpen
  \bibfield  {author} {\bibinfo {author} {\bibfnamefont {T.}~\bibnamefont {Senthil}}, \bibinfo {author} {\bibfnamefont {M.}~\bibnamefont {Vojta}}, \ and\ \bibinfo {author} {\bibfnamefont {S.}~\bibnamefont {Sachdev}},\ }\href {\doibase 10.1103/PhysRevB.69.035111} {\bibfield  {journal} {\bibinfo  {journal} {Phys. Rev. B}\ }\textbf {\bibinfo {volume} {69}},\ \bibinfo {pages} {035111} (\bibinfo {year} {2004})}\BibitemShut {NoStop}%
\bibitem [{\citenamefont {Guerci}\ \emph {et~al.}(2023)\citenamefont {Guerci}, \citenamefont {Wang}, \citenamefont {Zang}, \citenamefont {Cano}, \citenamefont {Pixley},\ and\ \citenamefont {Millis}}]{Guerci_SciAdv2023}%
  \BibitemOpen
  \bibfield  {author} {\bibinfo {author} {\bibfnamefont {D.}~\bibnamefont {Guerci}}, \bibinfo {author} {\bibfnamefont {J.}~\bibnamefont {Wang}}, \bibinfo {author} {\bibfnamefont {J.}~\bibnamefont {Zang}}, \bibinfo {author} {\bibfnamefont {J.}~\bibnamefont {Cano}}, \bibinfo {author} {\bibfnamefont {J.~H.}\ \bibnamefont {Pixley}}, \ and\ \bibinfo {author} {\bibfnamefont {A.}~\bibnamefont {Millis}},\ }\href {\doibase 10.1126/sciadv.ade7701} {\bibfield  {journal} {\bibinfo  {journal} {Science Advances}\ }\textbf {\bibinfo {volume} {9}},\ \bibinfo {pages} {eade7701} (\bibinfo {year} {2023})}\BibitemShut {NoStop}%
\bibitem [{\citenamefont {Zhao}\ \emph {et~al.}(2023)\citenamefont {Zhao}, \citenamefont {Shen}, \citenamefont {Tao}, \citenamefont {Han}, \citenamefont {Kang}, \citenamefont {Watanabe}, \citenamefont {Taniguchi}, \citenamefont {Mak},\ and\ \citenamefont {Shan}}]{Zhao_Nature2023}%
  \BibitemOpen
  \bibfield  {author} {\bibinfo {author} {\bibfnamefont {W.}~\bibnamefont {Zhao}}, \bibinfo {author} {\bibfnamefont {B.}~\bibnamefont {Shen}}, \bibinfo {author} {\bibfnamefont {Z.}~\bibnamefont {Tao}}, \bibinfo {author} {\bibfnamefont {Z.}~\bibnamefont {Han}}, \bibinfo {author} {\bibfnamefont {K.}~\bibnamefont {Kang}}, \bibinfo {author} {\bibfnamefont {K.}~\bibnamefont {Watanabe}}, \bibinfo {author} {\bibfnamefont {T.}~\bibnamefont {Taniguchi}}, \bibinfo {author} {\bibfnamefont {K.~F.}\ \bibnamefont {Mak}}, \ and\ \bibinfo {author} {\bibfnamefont {J.}~\bibnamefont {Shan}},\ }\href {\doibase 10.1038/s41586-023-05800-7} {\bibfield  {journal} {\bibinfo  {journal} {Nature}\ }\textbf {\bibinfo {volume} {616}},\ \bibinfo {pages} {61} (\bibinfo {year} {2023})}\BibitemShut {NoStop}%
\end{thebibliography}%
\end{document}